The risk factors affecting to the software quality failures in

Sri Lankan Software industry

By

Namadawa Bashini Jeewanthi Gamage

MBA                                                                    2017

The risk factors affecting to the software quality failures in

Sri Lankan Software industry

By

N. B. J. Gamage

A project submitted to the University of Sri Jayewardenepura in

partial fulfillment of the requirements for the Degree of Master of

Business Administration (Information Systems) on 31-01-2017

**Declaration of Candidate**

The work described in this thesis/research/project was carried out by me under the supervision of Dr. (Mrs.) S.M. Samarasinghe and a report on this has not been submitted in whole or in part to any university or any other institution for another Degree/ Diploma.

……………………………

……………………..

N. B. J. Gamage                                                    Date

**Declaration of Supervisor**

I/We certify that the above statement made by the candidate is true and that this project is suitable for submission to the University for the purpose of evaluation.

…………………………….

…………………….

Dr. (Mrs.) S.M. Samarasinghe                                        Date

Department of Information Technology,

University of Sri Jayawardenepura

# TABLE OF CONTENTS

## LISTS OF FIGURES

# LISTS OF TABLES

**ABBREVIATIONS**

| | |
|---|---|
| QA | : Quality Assurance |
| USD | : United States Dollar |
| USA | : United States of America |
| IT | : Information Technology |
| UK | : United Kindom |
| PMBOK | : Project Management Book Of Knowledge |
| IEEE | : Institute of Electrical and Electronics Engineers |
| OSS | : Open Source Software |
| ATM | : Automated Teller Machine |
| POS | : Point Of Sale |
| GSM | : Global System for Mobile communication |
| GPS | : Global Positioning System |
| ERP | : Enterprise Resource Planning |
| SAP | : Systems, Applications and Products |
| PLC | : Public Limited Company |
| CRM | : Customer Relationship Management |
| CMMI | : Capability Maturity Model Integration |
| TQM | : Total Quality Management |
| RBS | : Risks Breakdown Structure |
| IS | : Information Systems |
| SLASSCOM | : Sri Lanka Association of Software and Service Companies |

**ACKNOWLEDGEMENTS**

I would like to express my deepest gratitude to many people for their continuous support and encouragement, without which this achievement would not have been possible. In particular, I am extremely grateful to my supervisors: Dr. (Mrs.) S.M. Samarasinghe of the Department of Information Technology of Faculty of management studies and commerce at Sri Jayawardhanapura University of Sri Lanka. I would like to thank her for his enthusiastic guidance, invaluable advice, immense patience and also encouragement when times were tough throughout my MBA journey.

Thanks to Mr. W.M.N. Fernando who guide me to conduct this research. Thanks also go to Dr. Nalin Asanka Gamagedara Arachchilage, who pushed and continuously encouraged me to start a MBA project journey in the first place.

Thanks also to my friends who support me to collect actual data by sharing experience and comments in Sri Lanka software industry.

I would like to acknowledge the academic staff, colleagues and friends in 2014 MBA batch of Sri Jayawardhanapura University for their invaluable assistance and advice contributed to this research.

Special thanks must go to my dearest parents Lalitha Gamage and Wijepala Gamage, who have always encouraged me in my academic pursuits, given me the self-confidence to take on new challenges and supported me all the time.

The risk factors affecting to the software quality failures in Sri Lankan software industry

By N. B. J. Gamage

**ABSTRACT**

Software project failure and cancellation rates increase day by day due to technical failures, quality failures, lack of end client acceptance etc. and also the lack of proper management. Software project failure and cancellation rates increase due to the fact they have not been managed properly. This eventually leads a financial loss to the organization, perhaps the particular organization may require to obtain legal assistance, when the customer is not happy due to the breach of the contract. There are a number of reasons affected by the software project failures. According to empirical evidence, inadequate testing resources are one of the major factors that contribute to the poor quality.

The main objectives of this study are to study the risk factors that affect the software quality to provide some recommendation to minimize the risk of poor quality. There are three main factors affecting to software quality namely proper testing, test planning and QA team which are directly impacted to the software quality risks. To conduct this study, I employed an open-ended questionnaire for collecting qualitative data from responses analyzed them using thematic approach method.

The participants with their experiences agreed only with requirement clarity and clearly defined acceptance criteria, not with adequate unit testing and finally and also with that not doing regression testing force to quality failures. As of data analysis, not having proper formal test planning, initial test planning not being realistic, not following quality risk management, non-proper process and contingency action planning also lead to the risk of poor project quality. According to the participants added that the following factors are also behind the reasons for the lack quality of software. The experienced and skilled employees move out from the company as there is not a proper QA process and team members as they do not have the risk management mentality.

# CHAPTER 1: INTRODUCTION

## 1.1 Background of the study

Software project failures are rising up every day (Emam & Koru, 2008). Whittaker's 'What went wrong? Unsuccessful information technology projects' article (1999) states that business world invests billions of US dollars for information technology systems every year. As an example, in the USA, clients spend nearly 250 billion dollars for software developing in each year (Whittaker, 1999). They spend money on IT hardware, software and software after services, software upgrades, data wear housing services and license fees (Charette, 2005). They expect to earn the money back through return profit and benefits that provided by the systems (Costa, Barros & Travassos, 2007). Research failure costs are increasing day-by-day (Charette, 2005). As an example in 2005, Hudson Bay Co, Canada losses $33.3 million due to an inventory system problem and UK Inland revenue tax credit overpaid $3.45 billion due to software errors (Charette, 2005). These kind of errors and failures cause profit loss or project cancellations in organizations. Due to software project failures and cancellations, customers will have to bear high financial losses.

Analyzing and understanding software cancellation rates or failure rates could help to identify benchmark of the firm performance in the IT industry (Emam & Koru, 2008). Gathering reasons for software project quality failures is the aim of this research study. There are a number of reasons for software project failures. The most common reasons for software project failures are poor project planning, weak business cases and lack of top management involvement and support (Whittaker, 1999). The successfulness of a project can be measured in three different ways. They are project team evaluation, quality of project deliverables and client assessment (Pinto & Mantel, 1990). Rajkumar and Alagarsamy (2013) state that lack of testing resources leads to poor quality. Poor product quality is a major reason for project failures. Software failure due to the lack of quality is one of the risks of a project.

Carrying out a risk identification process is important before starting the project. Mc Connell (1997) state that when the risks are clearly identified and managed on time, there is a 50%-70% chance for a project become successful. However, the project budget increases by 5% with additional task of risk management (Kumar & Yadav, 2015). In risk management, software risks can be divided into two categories, systemic software risks and specific software risks. Systematic software risks are risks that directly affect software performance while specific software risks are risks that might hit to its success (Costa et al., 2007). Testing coverage could be the source of systematic risks and poor software quality might cause specific risks. Testing the product better will increase the product quality and in turn the project success rate.

Risk management of a software project is cost effective and it directs to long run sustainability. According to Knox 'reducing quality costs is to invest in defect prevention processes' (1993). Considering the quality and quality associated risk factors more could minimize the cost of software development and maintenance. As a guide to the project management body of knowledge (2013) state, to avoid risks related to quality, one needs to follow following steps. They are establish to the context, identify the risks, analyze the risks, evaluate the risks, treat the risks, monitor and review and communicate and consult (PMBOK, 2013). Furthermore, quality mangers and project managers could decide whether the identified risks could be avoided, reduced, transferred or retained. According to the Baccarini, Salm and Love (2004) 'continuous changes to requirements by the client', 'unrealistic expectations', and 'incomplete requirements' highly impact the project quality and scope risks. 'Poor production system performance' has an impact on both project and process management (Baccarini et al., 2004). There are a lot of researches and case studies for project failure risk factors identification. These researches study software quality related factors in general. However, there is no study conducted to deeply analyze risk factors affecting software project failures due to quality. To fulfill the gap between industry practices and researcher empirical findings, it is supposed to conduct this study based on Sri Lankan software industry.

Main objective of this project is to study the risk factors affecting the software quality and to give some recommendations on minimizing the risks related to quality.

## 1.2 Problem statement

Software project failure and cancellation rates increase day-by-day (Emam & Koru, 2008). 11% of software projects are canceled even before they are delivered due to the critical quality problems with the software (Emam & Koru, 2008). There are a number of factors affecting software failures. One of the main reasons for software failures is the lack of quality (Baccarini, Salm & Love, 2004). Therefore, poor quality could be due to factors such as, poor knowledge of quality assurance teams, poor application performance quality and delivered document quality problems, and frequent change of requirements (Pinto & Mantel, 1990).

Both software project failure and cancellation due to quality are risks to the firm's sustainability and quality assurance team performance. There are less research findings for quality assurance process at micro level analysis related to risk factors of quality.

However, there is a research gap between empirical findings and real world practice about the quality failure affected risk factors in QA team and process level.

## 1.3 Research question

What are the risk factors affecting to the software quality failures?

## 1.4 Objectives

Objectives of this study are;

1. to identify the risk factors that affect on the software quality.

2.  to give some recommendations on minimizing the risk of poor software quality.

## 1.5 Significance

Companies use quality as a weapon (Cope, Folse & Cope, 1999). Improving quality and avoiding or mitigating risks of quality is important. The reasons of lack of quality software and software project failure due to not having proper risk management (Kumar & Yadav, 2015). Risk management focuses on the avoidance of loss from unexpected events (Williams, 1995). This study helps the quality managers and leads to take some necessary actions to manage the identified risks. They will be able to understand how much testing; test planning and QA team skills affect the software product quality. According to the impact of each variable, they could decide on how much of attention should be paid on each and every variable from the total time of the project and the budget. They should be able to create or maintain a process to improve the quality of the project deliverables.

According to Bannerman (2008), "managers tend not to accept risk estimates given to them because they see risk as subject to control. They believe that risks can be reduced or dissolved by using their managerial skills to control the dangers". Managers always focus to alternative solutions for risk management than accepting and finding actual solution (March & Shapira, 1987). The results of this study provide them with actual risk factors that affect software quality. Quality managers can easily find solutions for factors affecting the quality.

## 1.6 Scope of the study

This research project is conducted to study the risk factors affecting the software quality. This study is conducted under three main categories. They are quality of testing, test planning and quality assurance team. In this study, a data collection survey is conducted, among Software quality managers and quality leads of Software development companies in the Sri Lanka. This study will fully focus on quality assurance process.

## 1.7 Chapter organization

Chapter 1 discusses the background of the research and the research gap of the existing knowledge and research question. Main objective of the research is to find out factors affecting the software quality risks. Benefited parties and people are boundaries of this study.

Chapter 2 is devoted to discuss existing literature to investigate why software projects fail or their cancellations occur, how software projects fail or their cancellations occur due to the lack of quality, the main risks factors affecting the software project failure? The empirical findings, clearly figure out research conducting methodology and population of the study. Supposed data sample collection is from targeted participants. Qualitative data analysis approach is selected according to the collected data type.

Chapter 3 is devoted to discuss and analyze collected qualitative data sample by using thematic approach and also discuss findings with other existing empirical research findings and participants' experiences related to the software quality risks.

Chapter 4 describes the research methodology, population of the study and the sample. It also includes a discussion of approach used to data collection and analysis.

Chapters 5 conclude all the findings with project risk management theories and earlier research knowledge. According to the study findings, it suggests some recommendations to minimize software project failures or cancellation due to the lack of quality.

# CHAPTER 2: LITERATURE REVIEW

## 2.1 Introduction

This chapter contains literature related to the Software project failure or cancellation, including the causes behind them and project quality failure risks. It also defines software quality and its importance to the software project's sustainability. If software project fails or is canceled due to the quality failure, it suggests that these are factors affecting software quality failure. When identified these factors turn in to project risk in last stage. It helps the main responsible parties to identify, the project quality failures. Main objective of this study is to figure out existing knowledge about software project quality failure risks.

## 2.2 Software projects failure or cancellation

The projects started before the industrial revolution and consequently people tend to do team work, even in their day today work when resources became widely available in the world (Mooney, 2011). They use resources for continuing their daily livelihood, such as farming, fishing, hunting etc. After the industrial revolution, machineries changed to life style of the man. People started to use equipments to save time and to make their work easy. With the change of life style of the man resources like human capital, power, money and time become scarce resources (Boehm, 1984).

2.2.1 Why people move to information systems/ software usage?

People move to Information Communication Technological systems because they need to increase their living standards (Hammond & Hammond, 1917). They need to save their time, money and energy as much as possible (Mooney, 2011). Handling goods at ware houses is much expensive task with a large inventory of stocks. Therefore, moving to computer based systems and payroll systems are cost saving actions (Charette, 2005). In 1950 electronic computer was produced and with the invention of internet and its impact on culture, commerce, and technology, the world changed to a global village in 1980. Then people started sharing their knowledge, data and information through information systems quickly, accurately and safely. Nielsen, Mack, Bergendorff & Grischkowsky (1986, p. 162) said, "Professionals working in a heterogeneous software environment are filled with practical problems, they follow "satisficing" strategies of sub-optimal usage, and they have problems migrating to more advanced uses. Current levels of software integration do not always adequately or easily support the "task integration" requirements of realistic tasks such as handling many small things."

## 2.2.2 Main software projects

In the software industry, there are two main types of software projects, development projects and maintenance projects. Software development projects are mainly a team, providing automated solutions for company activities, and providing necessary information for decision making with different type of formats (López & Salmeron, 2012). A software maintenance project focuses on development of some critical issues in the life-cycle of an enterprise system applications. Software maintenance projects do fixing bugs, improving performances and changing requirement enhancements (López & Salmeron, 2012). There are different types of maintenance projects such as, Corrective maintenance projects: These projects are fix design bugs, code bugs and functionality bugs. Adaptive maintenance projects, are develop a new environment, adapting user requirement changes. Perfective maintenance projects, improve application performance, cost effectiveness, efficiency and maintainability (López & Salmeron, 2012). Preventive maintenance projects, through many rounds of testing, find critical problems from the existing system that carry risks. They are fixing these critical issues (Burch & Grupe, 1993). User support projects, activities on time usage problems and user request supports and user training needs (Abran & Nguyenkim, 1991).

## 2.2.3 Open source software and Closed source software

IEEE definition of software is computer programs, procedures, and possibly associated documentation and data pertaining to the operation of a computer system (Standing Coordinating Committee of the Computer Society of the IEEE, 1990). Open source software (OSS) is "Software which has ability to distribute freely with available source code through the internet and using unpaid people that can modify the code freely" (Gacek & Arief, 2004). Also, Jayawarna and Fonseka (2011) defined computer software as software for which the source code and certain other rights normally reserved for copyright holders are provided under a software license that meets the open source definition or that is in the public domain. Some examples are OSS area, Linux, GNOME, Apache, Firefox, dove-cot, Open Office, Moodle and MySQL etc. It should have the many characteristics such as distributed software, free software, source code available, developers communicate through the internet, developers are users and unpaid and large amount volunteers (Gacek & Arief, 2004).

Jayawarna and Fonseka (2011) 'Closed source' is a term for software whose license does not allow for the release or distribution of the software's source code. Generally, it means only the binaries of a computer program are distributed and the license provides no access to the program's source code (Jayawarna & Fonseka, 2011). Examples for license soft wares are the Microsoft Windows, Macromedia Flash, Macromedia Dreamweaver, Adobe Photoshop, Mac OS, WinZip, Oracle, etc.

### 2.2.4 Software project incurred cost

Now software are used everywhere as an example, to get cash from ATM machines (Automated Teller Machine) and POS (Point-Of-Sale) machines in every shopping centers, and to make calls GSM (Global System for Mobile communication) technology with latest mobile phones available. People use somewhere GPS (Global Positioning System) available in this global village (Charette, 2005).

### 2.2.5 Software Direct cost

Software direct costs are main software development process related costs. Most countries' government bodies and organizations spent one trillion US Dollars for IT hardware, software, and services. United Kingdom government spent $ 20.3 billion in 100 IT projects in 2003. The United Stated government spent $ 16 billion only for military software systems (Charette, 2005).

### 2.2.6 Software Indirect/Overhead cost

Software indirect or overhead cost means, cost incurred for none-directly related to the development of a project. Currently, most commonly using, popular and leading ERP (Enterprise resource planning) application is SAP. But it is not user friendly to use. It needs more expert knowledge for integrating SAP systems to excising information systems and users training for use SAP system etc. In indirect/ overhead cost (List some disadvantages of SAP, 2016).

### 2.2.7 Software failure cost

Software failure incurred cost in 2004 October British food retailer J Sainsbury PLC installed a new automated supply chain management software to increase their business productivity for example. They invested US $526 million for the project. But they have

to hire 3000 extra clerks to do stock updates manually, because after implementing the automated supply chain system, some portals were stuck and failed to work successfully (Charette, 2005).

2.2.8 Software project failure situations in world history

Defining project failure is the most difficult task. Project failure depends on three incidents. They are the way in which failure is defined, the type of project being studied; and the stage of the project's life cycle at the time it is assessed (Pinto & Mantel, 1990).

These are some few examples listed in Table 2.1 about World history project failure and their value. It explains impact of project failure and its level of contribution it has done to the company's financial loss.

Table 2.1: Project failure and their value

| Year | Company | Outcome (Cost in US $) |
|------|---------|------------------------|
| 2005 | Hudson Bay Co. [Canada] | Problems with inventory system contribute to $33.3 million* loss. |
| 2004-05 | UK Inland Revenue | Software errors contribute to $3.45 billion* tax-credit overpayment. |
| 2004 | Avis Europe PLC [UK] | Enterprise resource planning (ERP) system canceled after $54.5 million† is spent. |
| 2004 | Ford Motor Co. | Purchasing system abandoned after deployment costing approximately $400 million. |
| 2004 | J Sainsbury PLC [UK] | Supply-chain management system abandoned after deployment costing $527 million. † |
| 2004 | Hewlett-Packard Co. | Problems with ERP system contribute to $160 million loss. |
| 2003–04 | AT&T Wireless | Customer relations management (CRM) upgrades, problems lead to a revenue loss of $100 million. |
| 2002 | McDonald's Corp. | The Innovate information-purchasing system canceled after $170 million is spent. |
| 2002 | Sydney Water Corp. [Australia] | Billing system canceled after $33.2 million† is spent. |

| 2002 | CIGNA Corp. | Problems with CRM system contribute to $445 million loss. |
|------|-------------|-----------------------------------------------------------|
| 2001 | Nike, Inc. | Problems with supply-chain management system contribute to $100 million loss. |
| 2001 | Kmart Corp. | Supply-chain management system canceled after $130 million is spent. |
| 2000 | Washington, D.C. | City payroll system abandoned after deployment costing $25 million. |
| 1999 | United Way | Administrative processing system canceled after $12 million is spent. |
| 1999 | State of Mississippi | Tax system canceled after $11.2 million is spent; state receives $185 million damages. |
| 1999 | Hershey Foods Corp. | Problems with ERP system contribute to $151 million loss. |
| 1998 | Snap-on Inc. | Problems with order-entry system contribute to a revenue loss of $50 million. |
| 1997 | U.S. Internal Revenue Service | Tax modernization effort, canceled after $4 billion is spent. |
| 1997 | State of Washington | Department of Motor Vehicle (DMV) system canceled after $40 million is spent. |
| 1997 | Oxford Health Plans, Inc. | Billing and claims system problems contribute to quarterly loss; stock plummets, |
| 1996 | Arianespace [France] | Leading to $3.4 billion loss in corporate value. |
| 1996 | FoxMeyer Drug Co. | Software specification and design errors cause $350 million Ariane 5 rocket to explode. |
| 1995 | Toronto Stock Exchange [Canada] | $40 million ERP system abandoned after deployment, forcing companies into bankruptcy. |
| 1994 | U.S. Federal Aviation Administration | Electronic trading system canceled after $25.5 million** is spent. |
| 1994 | State of California | The Advanced Automation System canceled after $2.6 billion is spent. |
| 1994 | Chemical Bank | DMV system canceled after $44 million is spent. |
| 1993 | London Stock Exchange [UK] | The software error causes a total of $15 million to be deducted from 100 000 customer accounts. |
| 1993 | Allstate Insurance Co. | Taurus stock settlement system canceled after $600 million** is spent. |
| 1993 | London Ambulance Service [UK] | Office automation system abandoned after deployment, costing $130 million. |
| 1993 | Greyhound Lines, Inc. | Dispatch system canceled in 1990 at $11.25 million**; second attempt abandoned after revenue loss of $61 million. |

| 1992 | Budget Rent-A-Car, Hilton Hotels, Marriott International, and AMR [American Airlines] | Travel reservation system canceled after $165 million is spent. |
|------|------|------|

\* Converted to U.S. dollars using current exchange rates as of press time.

† Converted to U.S. dollars using exchange rates for the year cited, according to the International Trade Administration, U.S. Department of Commerce.

\*\* Converted to U.S. dollars using exchange rates for the year cited, according to the Statistical Abstract of the United States, 1996.

Most of the companies allocate nearly 4%- 5% annual revenue for their information technology (Charette, 2005). If they invested projects get failed it could damage their company growth.

Sources: Business Week, CEO Magazine, Computerworld, Info Week, Fortune, The New York Times, Time, and The Wall Street Journal (adapted from Charette (2005))

## 2.2.9 Reasons of software project failures and cancellation

As the English idiom "Prevention is better than cure" suggests, some of investors decide to cancel ongoing projects before they fail, because they want to minimize the damage of failure. Projects fail because they are unable to achieve or partially achieve time, scope and quality requirements as end customer expected (Rajkumar & Alagarsamy, 2013). But cancellation of a project is wasting cooperate resources and it reflects poor management skills of the project managers (Emam & Koru, 2008).

## 2.2.10 Reasons for software project failures

IT project failures can also impact the economic growth of the country and quality of life of the citizens (Charette, 2005). Reasons for project failures are lack of customer or user involvement, unclear goals and objectives, poor requirement set, lack of resources, failure to communicate and act as a team, project planning and scheduling, cost estimation, inappropriate estimation methodology, cost estimation tools, poor testing, risk management, unrealistic expectations, poor reporting of the project's status, use of immature technology, inability to handle the project's complexity, sloppy development practices, poor project management, stakeholder politics and commercial pressures (Charette, 2005; Rajkumar & Alagarsamy, 2013).

## 2.2.11 Reasons for software project Cancellations

Software Product managers or a client should be able to cancel ongoing projects at any time of the software life cycle. As an example, 15.52 percent of projects were cancelled in 2005 and 11.54 percent were cancelled in 2007, before any delivery was made (Emam & Koru, 2008). Table 2.2 shows statistical details of project cancellations with percentage and 95% confidence intervals.

Table 2.2: Statistical details with percentage and reason for cancellation

| Reason for cancellation | Percentage of respondents (95% confidence interval) |
|---|---|
| Senior management not sufficiently involved | 33% (13, 59) |
| Too many requirements and scope changes | 33% (13, 59) |
| Lack of necessary management skills | 28% (10, 54) |
| Over budget | 28% (10, 54) |
| Lack of necessary technical skills | 22% (6, 48) |
| No more need for the system to be developed | 22% (6, 48) |
| Over schedule | 17% (4, 41) |
| Technology too new; didn't work as expected | 17% (4, 41) |
| Insufficient staff | 11% (1, 35) |
| Critical to quality problems with software | 11% (1, 35) |
| End users not sufficiently involved | 6% (0, 27) |

Note: Reasons for project cancellation with percentages and 95% confidence intervals for the 2007 respondents (n = 18) * (Sources: adapted from Emam & Koru (2008))

## 2.2.12 Reasons for software errors

Software errors are the cause for poor software quality (Shalloway, 2016). It is important to investigate the cause of these errors in order to prevent them. Those are faulty definition of requirement, client – developer communication failures, deliberate deviation from software requirements, logical design errors, coding errors, non – compliance with documentation and coding instructions, shortcomings of the testing process, procedural errors and documentation errors (Galin, 2009)

## 2.2.13 Project success or failure measurements or indicates

The project success or failure could be defined in three different measurements (O'Brochta & Michael, 2002). The first measurement is internal process performance. That is to project team performances evaluation, skill gap analysis, meet project deadlines, meet technical goals of the project, and meet estimated budget and time, and team working (Pinto & Mantel, 1990).

The second measurement is project deliverables, quality and process quality (Pinto & Mantel, 1990). Project deliverables are developed application, application related documentation like release notes, user manuals, project plan documents and project agreement document etc.

Third measurement is client satisfaction assessment (Pinto & Mantel, 1990). This is only one aspect of external factors of the project success or failure measurement. It could depend on non-functional requirements of delivering application like performance, security etc. (Hameed, & Arachchilage, 2016; Arachchilage, Namiluko, Martin, 2013).

## 2.3 Software quality

According to the Capability Maturity Model Integration (CMMI), quality assurance definition is "A planned and systematic means for assuring management that the defined standards, practices, procedures, and methods of the process are applied" (Chemuturi, 1950, p. 10). According to the expanded IEEE software quality assurance is "the systematic, planned set of actions necessary to provide adequate confidence that a software development or maintenance process conforms to established functional, technical requirements as well as the managerial requirements of keeping to schedules and operating within the budget" (Galin, 2009, p. 27). Software consists of abstract sets of rules that govern the creation, transfer, and transformation of data (Zmud, 1980). Also "A set of systematic activities, providing evidence of the ability of the software process to produce a software product that is fit to use" defined by Otte et al. (Bahamdain, 2015, p. 461). Juran was developed a quality trilogy as, quality planning is started by identifying customers and their needs, and then developing a product that meets those needs and optimizing the product so as to meet the organization's needs as well as the customers' needs. That is, quality starts with specifications and design. Quality improvement is defined a process that can produce the product, and then optimize the process. That is, quality depends on the process. Quality control is a test and proves that the process can successfully produce the product, and then implement the proven process in operation. (Chemuturi, 1950)

As Handbook of Software Quality Assurance states that the software quality is "Quality is the degree to which an object (entity) (e.g., Process, product, or service) satisfies a specified set of attributes or requirements". The two aspects need to fulfill software quality, the concept of attributes and the satisfaction or degree of attainment of the attributes (Schulmeyer, 2007, p. 6).

### 2.3.1 Total Quality Management (TQM)

TQM is a long-term strategic issue which is about continuous improvement in all areas of the organization's activities (Keogh, 1994). The three major components of TQM are:

(1) A quality assurance system,

(2) Quality tools and techniques (Keogh, 1994) and

(3) Teamwork (Mortiboys & Oakland, 1991).

According to British Standard 7850: Total Quality Management, states that "Total quality management assures maximum effectiveness and efficiency within an organization by putting in place processes and systems which will ensure that every aspect of its activity is aligned to satisfy all customer needs and other objectives without waste of effort and using the full potential of every person in the organization" (British Standards Institution, 1992)

### 2.3.2 Software quality aspects

The main task of the development team is to convert end customers' requirement or idea or need to software. There are three teams which are well concerned about software quality. They are end users, sponsors and development team. These three teams focus on three different quality aspects. As per figure 2.1, the end users are more concerned about functional quality, development team members are mainly concerned about structural quality and Sponsors are concerned about process quality (Chappell, 2013).

Functional quality is the quality of software that meets functional requirements expected by the end user.

Structural quality means the quality of software that meets non-functional requirements supporting the delivery of the functional requirements.

Process quality means quality of project outcomes.

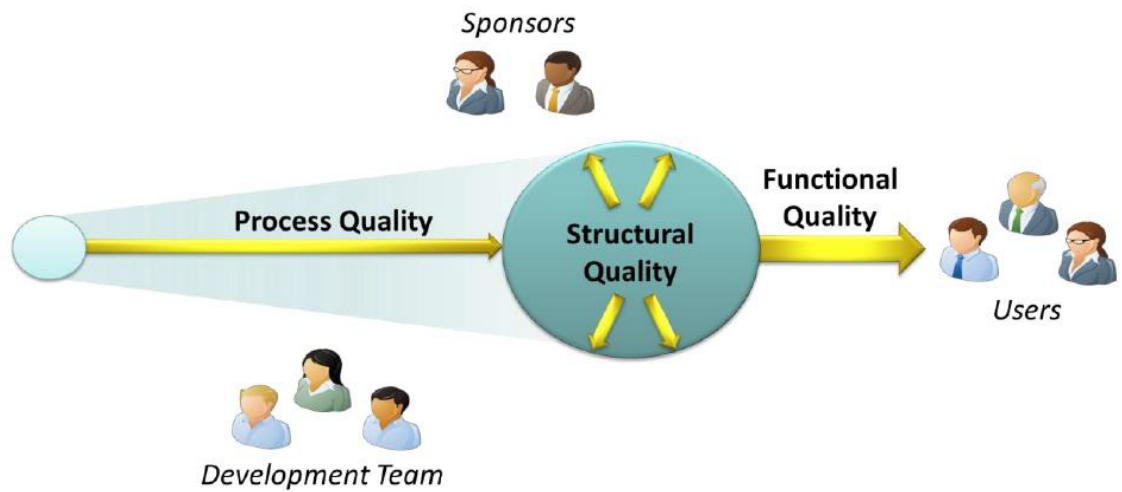

Figure 2.1: Software quality three aspects

Source: Software quality three aspects (adapted from Chappell, (2013))

2.3.3 Quality Policy

According to Regan (2002) the quality policy of the organization places responsibility for management, and employee and the quality policy needs to be actively implemented. This involves planning for quality, customer satisfaction, and continuous improvement. The quality policy needs to be periodically reviewed to ensure that it continues to meet the needs of the organization. Top management needs to ensure following quality policies, appropriate to the purpose of the organization, include a commitment to comply with requirements and continually improve the effectiveness of the quality management system and provide a framework for establishing and reviewing quality objectives, which are communicated and understood within the organization and are reviewed for continuing suitability (Galin, 2009).

### 2.3.4 Software quality Assurance

Accoring to Galin (2009) explains, the objectives of software quality assurance activities for software development and maintenance are assuring; With acceptable levels of confidence, conformance to functional technical requirements, assuring: with acceptable levels of confidence, conformance to managerial requirements of schedules and budgets and initiating and managing activities for the improvement and greater efficiency of software development and software quality assurance activities.

### 2.3.5 Software Testing

The main objective of software testing is to identify software faults and other failures to fulfill the requirements. Software tests examine software modules, software integration, or entire software packages (systems). Recurrent test (regression test) carried out after correction of previous test findings, are extended till satisfactory result are obtained. (Galin, 2009)

### 2.3.6 Software Planning

Top management shall ensure that quality objectives, product requirements are established at relevant function and level within the organization. Top management ensures that, the planning of the quality management system is carried out in order to meet the requirement as well as the quality objectives, and the integrity of the quality management system is maintained when changes to the quality management system are planned and implemented (Raju & Parthasarathy, 2009).

2.3.7 Software Tester

A team for performing system testing is truly separated from the development team, and it usually has a separated headcount and budget. Members of the system test group conduct different categories of tests, such as functionality, robustness, stress, load, scalability, reliability and performance (Naik & Tripathy, 2008)

## 2.4 Software quality related risks

Olds (1999) provided a systematic approach to planning and managing the software development process in a business context with an emphasis on advancing software quality while controlling business risk. The three components of Ould's approach are the business problem and its analysis, the risk and quality plan, and the project resource plan (Tsoukakas, J., 2001).

Quality Control tests prove that the process can successfully produce the product, and then the proven process is implemented in operation. (Chemuturi, 1950)

System test planning is to get ready and organized for test execution. A test plan provides a framework, scope, details of resource needed, effort required, schedule of activities, and budget (Naik & Tripathy, 2008).

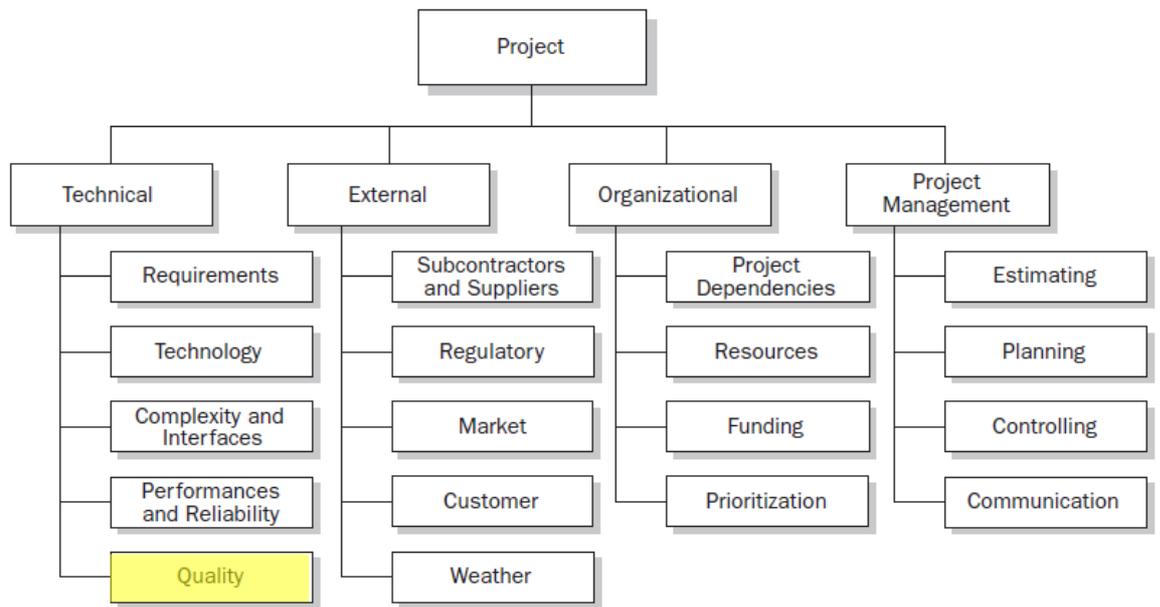

Figure 2.2: Risks Breakdown structure (RBS)

Source: Risk Breakdown structure (RBS) (adapted from PMBOK, (2013))

The software engineering work initially identifies the technical execution risk factors such as IT personnel skills, project size, technical complexity, cohesion of the project team, and a continuous stream of requirement changes (Boehm 1989). According to Figure 2.2 the Project Risk Breakdown Structure (RBS) Quality risk management plays a major role. Mainly project success is based on project quality. It could be code development, quality, application performance, quality, delivered document quality, after service (maintenance) quality or communication quality etc. There is a link between risk, flexibility, and real options (Benaroch, Lichtenstein & Robinson, 2006). Some work explains how three types of options (avoidance, reduction, transfer and retention) can help to justify certain project management decisions made in relation to certain IT risks. According to Zhi (1994), there are four main strategies for responding to project risks (Baccarini D., 2004).

(1) Avoidance is not undertaking the activity that gives rise to the risk.

(2) Reduction is reducing the probability of a risk event occurring, and/or the impact of that event. Risk reduction is the most common of all risk-handling strategies (Pritchard, 1997).

(3) Transfer is transfer of risk in whole or part to other party.

(4) Retention is accepting the risk and therefore the consequences it eventuates (Baccarini D., 2004).

According to the Costa et al. (2007, p. 17) "Project risk level is as the probability of a project failure in achieving its proposes goals". Thus, if a project has lower risk level (30%) it has a higher success chance (70%). Based on risk-and-return ratios, project managers easily assets and compare their projects according to the risk levels. (Costa et al., 2007, p. 17)

The software project risk could be classified into two categories, Systematic software risks and Specific software risks. Systematic software risks are mainly affected by the performance of the project and the specific software risk shows how much the project deviates from its chance of success (Costa et al., 2007). According to the behavioral view of risk, decision makers associate risk with a probability concept and with the degree of a bad outcome (March & Shapira, 1987).

Requirement changes, planning weaknesses and testing coverage could be classified as systematic software risks. Project development difficulties, lack of required hardware resources could be classified as specific software risk (Costa et al., 2007).

2.4.1 Software quality risk management importance

Software quality risk identification and listing are the most important steps before risk management, because every risk is bind with mitigation actions or resolve actions (Iversen et al., 2004). IT project managers have been observed to improperly evaluate risks before prioritizing them for management attention and thereby paying greater attention to some risks at the expense of others (Schmidt et al., 2001). Article Project risk management: lessons learned from software development environment show "Effective risk management is the most important management tool a project manager can employ to increase the likelihood of project success" (Kwak & Stoddard, 2004). Quality team risk management also a considerable factor. It helps to share risk responsibilities and burden effectively (Kwak & Stoddard, 2004). Furthermore, IT managers can adequate way to quantify the economic value of mitigations relative to risk (Benaroch et. al., 2006).

## 2.5 Available methodologies

In the analysis of software quality risks, the best method is qualitative approach. because it is interested in collecting critical aspects of software quality and risk management is a complicated process that is difficult to measure quantitatively (Patel, Mohanan, Prabhakaran & Huffman, 2016). According to the Bengtsson (2016), "Qualitative research contributes to an understanding of the human condition in different contexts and of a perceived situation". Therefore, in analyzing critical risk factors affecting the software quality research, the most suitable methodology is a qualitative approach than quantitative data analysis by statistical prediction. As figure 2.3 there are number of processed needs to be followed to qualitative research. However, Planning is important out of all the processes. According to planning, researcher needs to collect data, analyze data and finally produce a report to present the final results.

This chapter tries to elaborate planning process of this study. According to Downe - Wamboldt (1992), here it explains the process of study, about population and sample and their limitations.

In this study, the main aim is to identify risk factors affecting the software quality. The sample of the qualitative research should be 1 to 30 information (Fridlund & Hildingh, 2000). Also, the researcher should be guided by aim of the study. There are no specific rules for data collection method (Bengtsson, 2016). This case study is supposed to use open-ended questions as in a questionnaire (Donath, Winkler, Graessel, & Luttenberger, 2011). The choice of data collection method affects the depth of the analysis (Bengtsson, 2016). It is finally needed to choose a data analysis method. In qualitative study data analysis, data is presented in words and themes, which makes it possible to draw some interpretation of the results (Bengtsson, 2016).

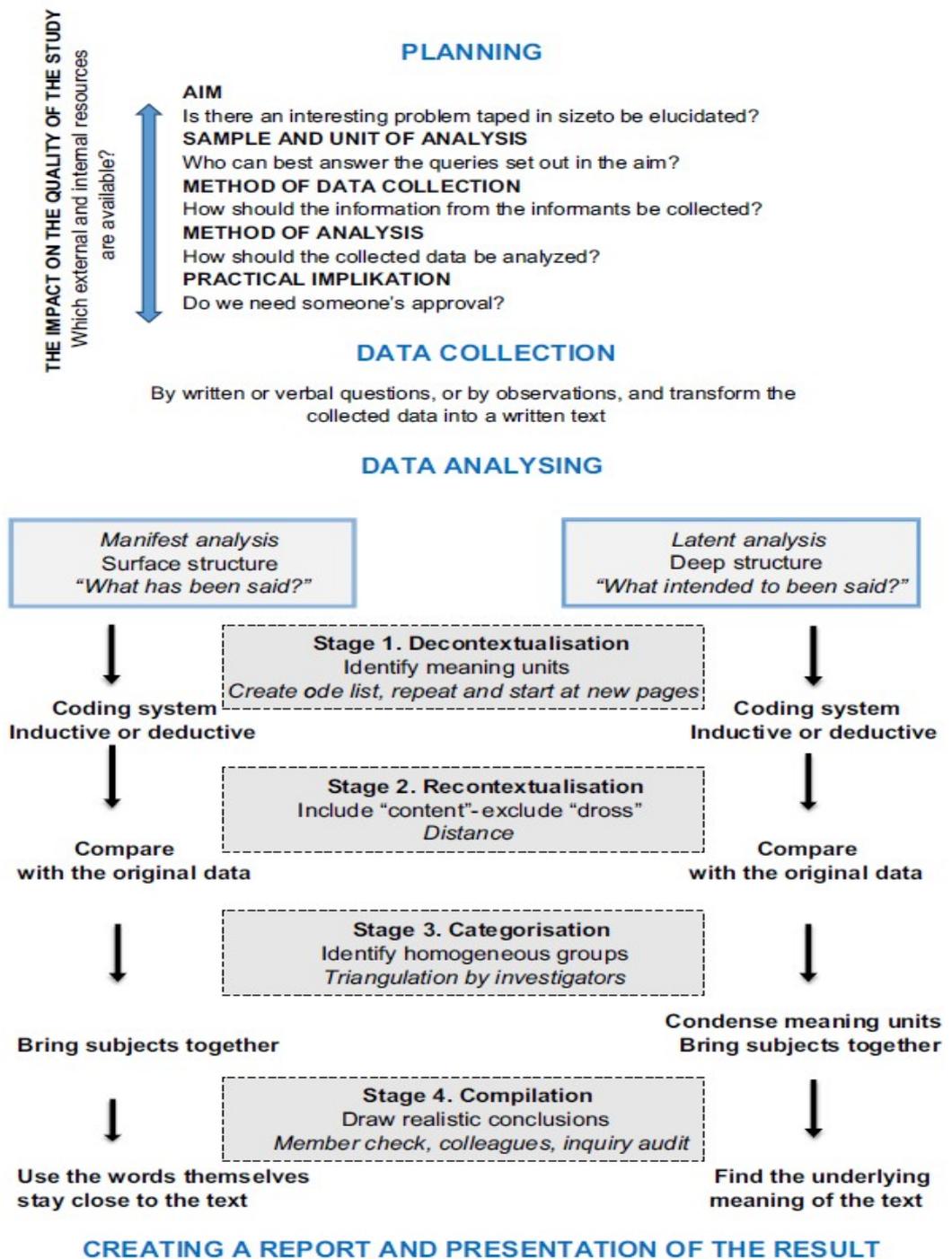

Figure 2.3: An overview of the process of a qualitative content analysis

Source: An overview of the process of a qualitative content analysis (adapted from Bengtsson, (2016))

## 2.6 Summery

This chapter provided the empirical evidence for software project failure or cancellation and reasons for them and factors affecting software failure and their incurred cost (Emam & Koru, 2008). Statistics shows that 11% of software projects fail due to the lack of quality. Furthermore, an analysis of factors affecting software quality risks is included. Risk management is an important process of software project management. In this chapter, available research findings and evidence related to the research gap are gathered. Finally, a broad description of qualitative research methodology and plan of the study is elaborated with empirical findings.

# CHAPTER 3: METHODOLOGY

## 3.1 Introduction

This chapter provides the research methods used to conduct the study. It explains how the necessary data and information are collected to address the research objectives. It also identifies research population and sample, the questioner and interview questions designed to fulfill research gap. It also give a discussion over the collected data presenting method and analyzing method.

## 3.2 Research Design

Main objective of this project is to study risk factors affecting to the software quality. According to research gap identification, targeted research area is software quality, and audience is software quality assurance professional's including project quality assurance managers, quality assurance leads, senior quality assurance engineers and quality assurance engineers. There are several research methodology approaches available in Information Systems literature"(Arachchilage, 2012; 2016; Arachchilage and Love, 2013; 2014; Arachchilage, 2015; Arachchilage and Martin, 2015; Arachchilage and Martin, 2013; Arachchilage and Asanka, 2012; 70. Arachchilage, Namiluko, Martin, 2013; Arachchilage, Love, & Beznosov, 2016). For example, the qualitative, the quantitative and the mixed method approach. According to the research methodology, qualitative approach could have been used in this research study. However, quantitative approach is selected because," it offers the flexibility to represent the population in general of users within organizations and also widely penetrated approach in IS" (Arachchilage and Love, 2013; Arachchilage and Love, 2014; Cherapau, Muslukhov, Asanka, & Beznosov, 2015) Furthermore, quantitative research approach can aid to use tables and charts to visualize the data, use appropriate means to describe it, and choose some methods to examine trend and relationship within it using statistical techniques. According to Patel et. al. (2016) software quality risk management process is a complicated process and it is complicated to measure quantitatively. Therefore, Patel et. al. recommended to follow a qualitative approach for analyzing software quality risk factors. To software projects, quality risk factors identification questions from questionnaire mentioned on 'Evaluating software project portfolio risks' articled by Costa et al. (2007). Therefore, exact software quality risk related questions are out of portfolio risk related questions. Also, the researcher conducted interviews with software quality managers of leading software companies in Sri Lanka.

### 3.3 Population and Sample

Population is the total collection of elements about which we wish to make some inferences (Cooper et. al., 2012). The population of this research consists of all software quality assurance professionals' in Sri Lankan software industry. Most of the software development and support companies located in Colombo district.

For this study 21 data were collected from 21 companies normally ShipXpress, a GE Transportation Company, Allion Technologies (Pvt) Ltd., Aviation based company, Auxenta (Pvt) Ltd., Datacom Group Ltd, eBuilder Technology Centre Pvt Ltd., Information and Communication Technology Agency of Sri Lanka, Dialog Axiata PLC., Synapsys (Pvt) Ltd., Pearson Lanka Pvt. Ltd., Virtusa Polaris Pvt. Ltd., Inexis Consulting, Teknowledge Shared Services (Pvt) Ltd., Aeturnum Lanka (Pvt) Ltd., Adelco (Pvt) Ltd., xcendant (Pvt) Ltd and 99X Technology.

Sampling is an extract of some countable elements from the population. It justifies for the lower cost, greater accuracy of results, greater speed of data collection and availability of population elements. Population sampling approach is random sampling method. Random sample gives a true cross section of the population (Cooper et. al., 2012).

Participants represented both government software projects and private software firms. Here 14% of participants are Quality managers and 36% of participants playing their project role as quality team lead of their engagements. All the survey questions were direct and targeted questions to share their experience and advice to a focused area of the researcher.

### 3.4 Questionnaire Design

An open-ended questionnaire was used to gather responses apart from the preliminary interview. Refer to (APPENDIX I) for further details. The questionnaire is divided in to four main sections as shown in table 3.1. Section one has seven questions, capturing organizational demographics of the responder. Section two had five questions, to capture responder's personal experience of testing related risks. Section three had five questions, to capture organizations' software quality planning related risks. Section four was targeted to capture quality assurance team skills and human resourcing related risks.

Table 3.1: Risk factors and their number of questions

| Factor | No. of Questions |
|---|---|
| Responder's demography | 7 |
| Testing | 5 |
| Planning | 5 |
| Team | 5 |

For the interview use open-ended questions were used focusing on the overall software quality related risks.

### 3.5 Approach to data collection

The study was carried out using questionnaires and interviews methods.

### 3.5.1 Questionnaire method

The research used unstructured questions to gather necessary data. Open-ended questions were used to ensure that respondents' feelings are not disregarded and participants further explanations are made. This survey study conducted through the internet and mobile platform support application. So, participants experience the data collection process is fast and easy than a manual process. As the first step of the data collection, created an e-survey questioner by using selected focused questions and participant's background related questions. The questionnaires were emailed in person. Questionnaires were distributed after initial communication with the respondents to get consent. The respondents were given one week to answer the questionnaires (Pinto & Mantel, 1990).

### 3.5.2 Interview method

An interview is an interactive forum involving two people engaged in a conversation initiated and coordinated by the interviewer so as to get information specific to software quality risks of aspect. Telephone call interviews were carried out with QA managers of reputed software companies of Sri Lanka. All interviews were carried out prior to setting of appointments with the concerned respondents. The interviews had specified time limits of approximately 10 minutes. All interviews were carried out with the help of already prepared interview guide question papers (see APPENDIX II) and were recorded alongside the respective questions.

### 3.6 Approach to data analysis

The data analysis was carried out with thematic analysis approach as described by Braun and Clarke (2006). Thematic analysis approach is a method which can identify data by analyzing and reporting patterns (themes) from data. Advantage of thematic analysis has flexibility (Bruan & Clarke, 2006). Thematic analysis approach has 6 phases given in as table 3.2.

Table 4.2: Phases of thematic analysis

| Phase | Description of the process |
|---|---|
| 1. Familiarizing yourself with your data: | Transcribing data, reading and re-reading the data, noting down initial ideas. |
| 2. Generating initial codes: | Coding interesting features of the data in a systematic fashion across the entire data set, collating data relevant to each code. |
| 3. Searching for themes: | Collating codes into potential themes, gathering all data relevant to each potential theme. |
| 4. Reviewing themes: | Checking if the themes work in relation to the coded extracts (Level 1) and the entire data set (Level 2), generating a thematic 'map' of the analysis. |
| 5. Defining and naming themes: | Ongoing analysis to refine the specifics of each theme, and the overall story the analysis tells, generating clear definitions and names for each theme. |
| 6. Producing the report: | The final opportunity for analysis. Selection of vivid, compelling extract examples, final analysis of selected extracts, relating back of the analysis to the research question and literature, producing a scholarly report of the analysis. |

Source: Phases of thematic analysis (adapted from Bruan & Clarke (2006))

### 3.7 Summery

This chapter described the methodology of this study. It also explained the steps of research design, questionnaire designing and approach to data collection and analysis. With refers to the qualitative research data analysis as conducted by other researchers, it is intended to use the thematic analysis approach for software quality data findings.

# CHAPTER 4: DATA ANALYSIS AND DISCUSSION

## 4.1 Introduction

This chapter elaborates on the analysis of data and discusses the result. Firstly, the demographic data of participants was analyzed. It also used for the analysis by using graphical method, the qualitative data analysis method. Core of the data is analyzed by using thematic analysis approach.

## 4.2 Data Analysis and Discussion

### 4.2.1 Demographic data analysis

Data was analyzed by classifying several groups according to the demographic details of the participants. According to figure 4.1 the majority of participants were had postgraduate qualification

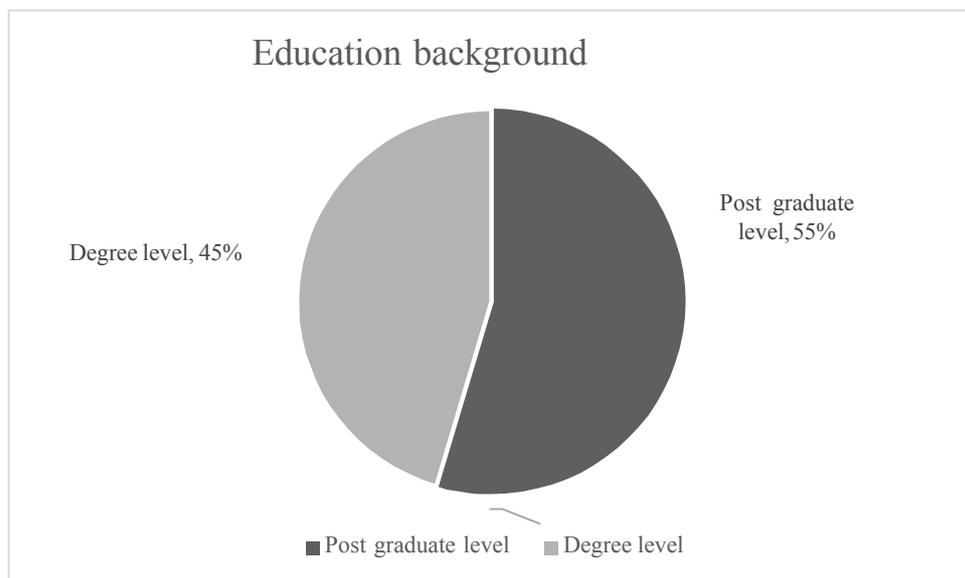

Figure 4.1: Education background

All QA managers participated in this study provided actual risk management issues and process related information. The rest represented the QA leads and senior QA engineers as in figure 4.2 below. They  shared their experiences and practices with this study.

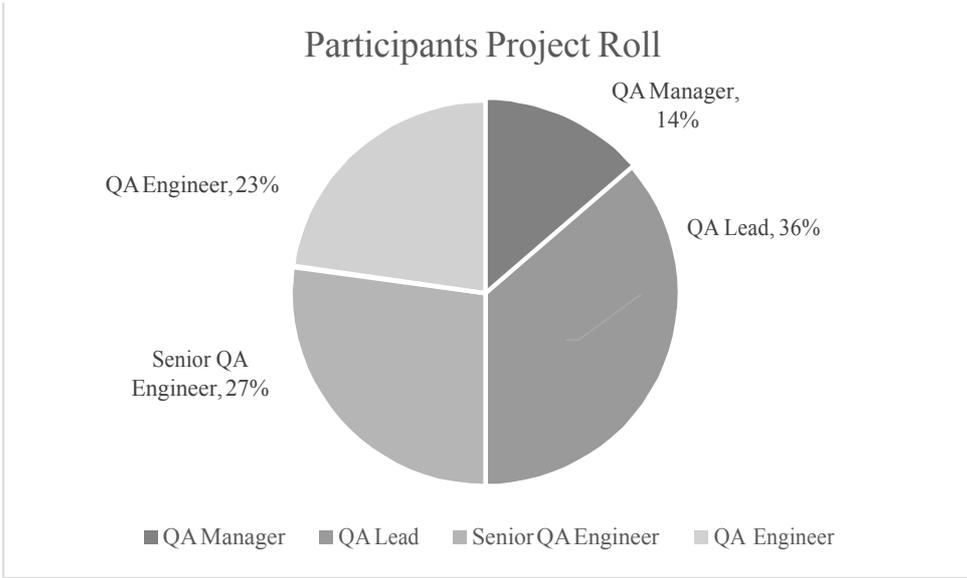

Figure 4.2: Project roll of respondents

The most important factor in this study is QA team size, because QA team size defines according to the project complexity and project size. Participants represented many different sizes of the QA teams on their companies as given in Figure 4.3. QA teams of most of the companies consisted of 3 or less than 3 QA members. That means they would not have enough team members to follow proper QA process.

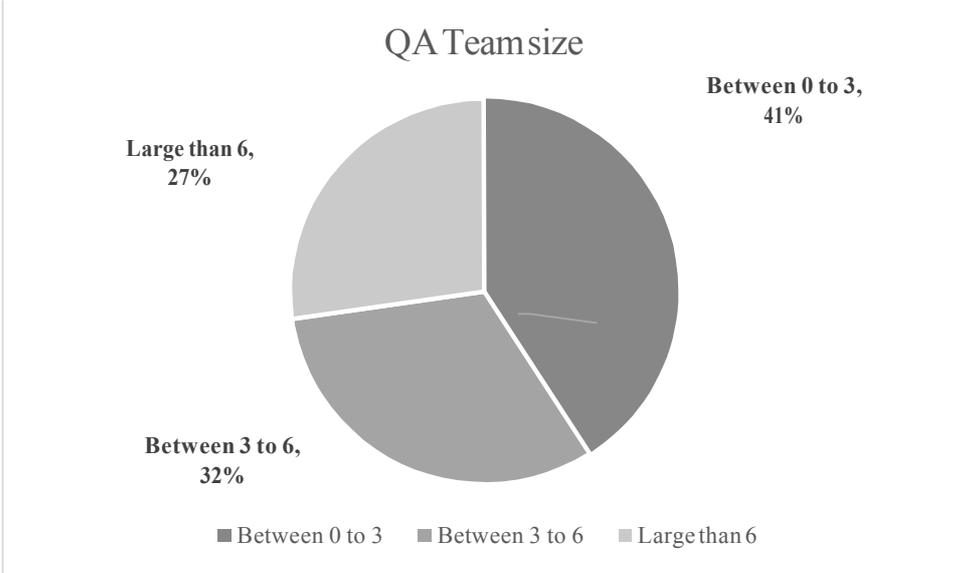

Figure 4.3: QA team size of responded

According to the figure 4.4 most of the companies are categorized as medium scale companies, as when they were in process development stage they need much experience or good consistency to establish proper project risk management process.

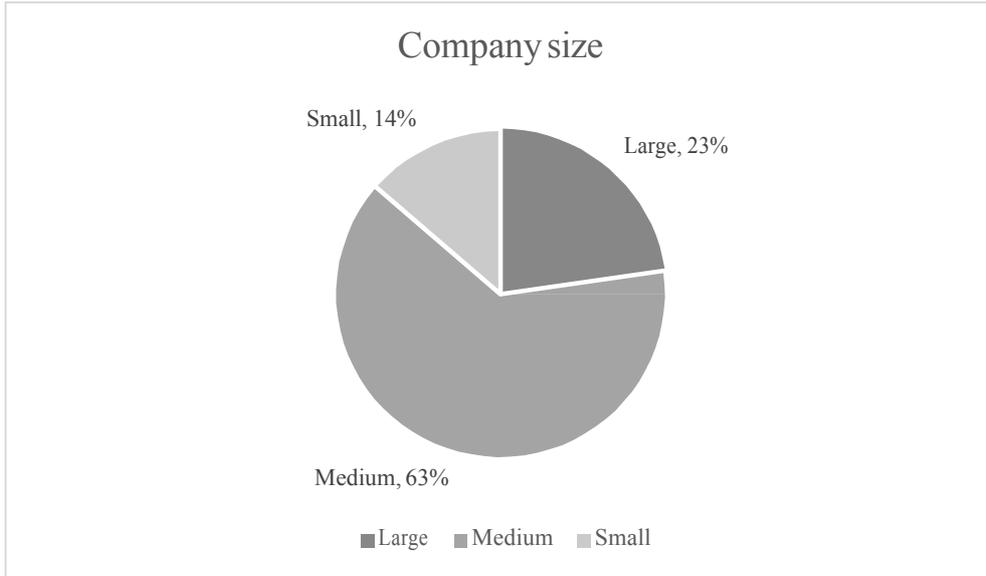

Figure 4.4: Company scale of responded

According to figure 4.5 most of the participants represented development projects and 23% represented maintenance projects. Therefore, there should be a proper project management process to avoid the risks of the project success.

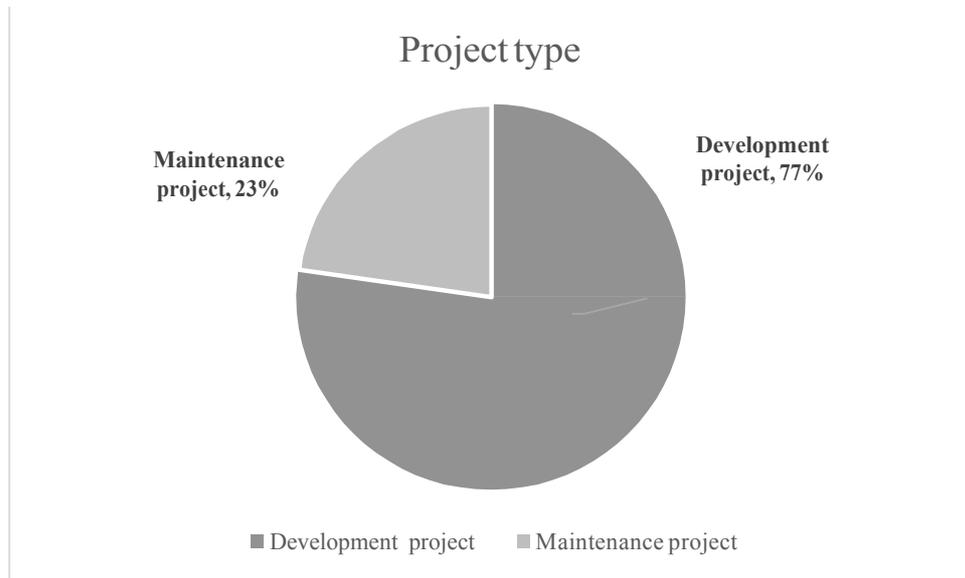

Figure 4.5: Project types of responded

## 4.2.2 Qualitative data Analysis

After following Thematic approach process steps, three different themes were identified,

Theme 1: Does not have enough testing leads to project quality risk.

Requirement clarity and clearly defined the acceptance criteria are the most important factors for doing testing correctly. Large companies always define acceptance criteria at the initial stage of the project. If there are any questions they would clarify them through business analyst agents of on site or offshore teams. They were strongly believed effective communication strong weapon of the project success.

*"Yes. Acceptance criteria agreed by the team and considered as meeting the definition of done."* (QA manager, Large scale company)

However medium and small scale companies' state that sometimes the client does not provide proper acceptance criteria on corrected time. Government attached software companies don't receive acceptance criteria every time.

Component wise unit testing gives a big impact to the software quality. Large companies always maintain checklists or task lists and ensure their test coverage and pass rates. Therefore, they could manage their final output quality in a good manner. But due to the lack of time QA teams of medium and small scale companies are unable to do enough unit testing. As an example the following quotations of two participant can be highlighted,

"*No. The development team and the QA team are unable to do sufficient unit testing due to strict timelines.*" (QA lead, Medium scale company)

"*It depends on the time given by the project manager. Sometimes in our company sufficient time is not given at all.*" (QA lead, Aviation based company)

Module integration testing is most value-added task of the QA life cycle. Therefore, QA team needs more time to perform integration testing. According to the large companies process, they take team participation for time estimation for system testing and integration testing in story point level. In exactly a one of large scale company QA manager's experience here,

"*System testing and integration testing is considered as per the need. The team agreed for the testing scope and provide estimates for testing during poker planning session. Further, effort estimate is considered after considering the story points.*" (QA manager, Large scale company)

However, small and medium scale companies also allocate time for the initial planning session and quality managers always encourage them to focus on integration testing.

Software companies should always support the team to produce quality products. For that they need to provide required hardware and software facilities. All the participants are happy with their companies support of hardware and software.

Regression testing plays a major role of the final product quality, because QA engineers could be identified impacted areas which from unplanned, bugs related code fixes. But with time considering most of the companies do not allocate time for regression testing. Refer to the, two quotes of participants from the large and medium scale companies given below,

*"As per the need, regression testing will be considered."* (QA Lead, Large scale company)

*"No, we don't have enough time allocated for regression testing."* (QA Engineer, Medium scale company)

Discussion of Theme 1:

In 2007 Costa, et al. did 'Evaluating software project portfolio risks' quantitative research study with 50 samples of participants. According to their analysis they give 10.81 weights for testing risk factor. Therefore, testing is effective 10% of all project portfolios. According to this, research integration tests and hardware and software facilities were not impacted to the software testing risk. Requirement clarity is important. The acceptance criteria should be clear to the reader (Koelsch, 2016) (both developers and testers), for application development as well as testing. Otherwise, functional development directs to an incorrect path and it could cause to quality failure as well as customer un-satisfying. According to Hamill (2005), "A single unit test should test a particular behavior within the production code." Unit test results are displaying the quality of end product development. As well as it displays the developers skills and performances. Running unit tests directly affect on the software quality improvement. Regression testing (as known as Confirmation testing) is verifying that changes in the software or the environment have not caused unintended adverse side effects and that the system still meets its requirements (Graham, Veenendaal & Evans, 2008). With project scope, size testing area is too large. Therefore, most of the time, automated test scripts are used for regression testing. That is time saving of final confirmation testing round. However, requires clarity, clear acceptance criteria, doing enough unit testing and final regression testing rounds are highly impacted to the software quality.

Theme 2: Does not do proper planning leads to project quality risk.

Planning is an important factor for quality management. Without proper planning and its executing managers could not track project status. All of the project quality teams do the formal of test planning. Following is the experiences from a large-scale company QA manager,

*"Yes. Test plan Covering risk, approach and testing types, schedules."* (QA manager, Large scale company)

Small scale company QA leads state that,

*"Yes, but sometimes it won't work out with emergency situations. With some hot fix releases, normal test plans cannot execute properly."* (QA lead, Small scale company)

However, sometimes those cannot execute properly due to some unexpected situations like requirement priority changes, the unexpected releases scheduled between main releases.

The project identified risk managing correctly leads to project success. Over 70% of the companies do not follow quality risk management proper process. Most of the time during the project testing phases, engineers identify risks. In a project, in daily standup meetings they come out with these risks. Therefore, project QA managers suggest necessary risk mitigation actions as given below,

*"No, we do not follow risk management process. As a QA team, we always highlighted the risk and the necessary actions will be taken to minimize the risk and proposed risk mitigation plans."* (QA lead, Medium scale company)

However, contingency action plan also created when there are any identified risks available or arise at that point. Large scale companies do follow past experience to create a contingency plan. But most of the medium and small scales have mainly the less number of projects, so sharing past experience is very rare they always try to find new contingency action plans. This process takes more time. Senior QA engineer shares her experience as;

*"Yes, if there are risk appears we discuss with QA management and find out contingency plans for them."* (Senior QA engineer, Medium scale company)

With the risks appears and other tasks, is QA team could have prepared plans and schedules realistic? According to the QA manager of a large company, they clearly align delivery plan with other tasks.

*"The team comes up with story points as per the decided scope of the requirements. If there are changes need to be done, then those will consider as new changes and will cater in future sprints with adequate time. Hence the schedule set is fine for delivering the minimal viable product."* (QA Manager, Large scale company)

Most of the QA team member states that they could not be achieve planned time limes due to some unexpected reasons. Such like below commented in medium scale company member.

*"QA schedules are realistic, but most of the time it's going out of the schedule due to the below; 1. Lack of time for the requirement analysis phase; 2. Issues in the development time estimations; 3. Ongoing risks"* (QA lead, Medium scale company)

All participants confirm that they do not have an issue with adequately team members. Most of the projects are agile project management. Therefore, managers assign resources as per project requirements and their priorities. As an example,

*"As per the project scope, resources are defined. As the team operates in an agile mode, resources are pulled in as per the requirement priorities."* (QA lead, Large scale company)

Discussion of Theme 2:

"Many projects encountered problems due to poor project setup" status by Bannerman, (2007) and Costa et al. (2007) weighted 13.51 for planning. Therefore, all researchers agree planning is an important factor of the project. Test plan explains other stakeholders how to testing accomplished. If the plan is developed carefully, test execution, analysis and reporting will flow smoothly (Maidasani, 2007). Some project planning did not align to value-adding business objectives or where the project setup was left to a dominant vendor whose priorities and actions were driven mainly by self-interest (Bannerman, 2007). These study participants also agree without proper formal test planning and quality risk management process, that project cannot deliver 100% quality warranty software. To manage quality up, and risk down, project planning and strategic decision making for software developers presents practical and realistic planning techniques to increase the chances of a project delivering to time and budget (Ould, 1999).

Real-time threat management capability is developed within an organization, through learning, practice, and other mechanisms, over a long period of time. Risk management is not just about identifying and assessing risks, and putting in place mitigation and contingency strategies (Bannerman, 2007). Also, they should have contingency action planning for real time appeared risks and prepared plans and schedules should drive to success.

Theme 3: Does not with proper QA Team resource lead to project quality risk.

As per large scale companies provide on-board training and when, after assigning to the projects, project seniors give them to domain knowledge training. If QA team members requests special training company provide trainings in a timely manner.

*"Yes, we recruit skilled people; the company gives domain training and other required trainings."* (Senior QA Engineer, Medium scale company)

Small and medium companies have a risk, because experienced and skilled employees are moving out from the company. But still large scale companies do not have such kind of risk.

*"Yes. With high stress of project management engineers tend to move out of the company."* (QA lead, Small scale company)

*"No, as a reputed company, our employee turnover is very low."* (QA lead, Large scale company.)

Currently, there is no issue of the QA team, because of their effective and efficient work. All participants are satisfied about team work towards one goal.

Following the QA process is one good option for minimizing project risk. Large and medium scale companies always follow and try to stick with QA process.

*"The team should meet the acceptance criteria to consider that definition of done is met. Hence team should work as per the set processes."* (QA engineer, Large scale company)

As of participants experience not all team members aware about risks and risk management. Only quality management level and senior level team members aware and actively participate risk management process.

*"Mostly in managerial level. but team will also provide their input during the reviews and retrospective sessions as per their experience."* (QA manager, Large scale company)

*"Yes. Seniors always monitor and concern with current risks and future risks."* (Senior QA Engineer, Medium scale company)

As a conclusion of data analysis lack of requirement clarity, unclearly define acceptance criteria, lack of unit testing and not performing regression testing can take as factors affected to the test coverage risk. Not having proper formal test planning, not following risk management process, not having a proper contingency plan and not achieving initial test schedules leads to planning risk. Experienced and skilled team members left out of the project, does not follow the proper QA process and every team member does not aware about risk management process leads to lack of required resources risk.

Discussion of Theme 3:

According to data given by participants are rejected the training factor and working for common goal factor. However, even in organizations with the best processes, skills, and organizations that motivate team members towards effective risk management, the uncertainties resulting from the sheer magnitude of software project complexities can make managing risk a daunting task, because of the imperfections of human judgment (Kwak & Stoddard, 2004). If companies do not motivate their skilled employee force, they will move out of the firm. It could be affected by the software quality. If the QA team does not follow the proper process, they could avoid mistakenly important steps of the quality assurance path. If team avoids or hides quality related risks, project might move to unsuccessful and failure.

Open ended question gives freedom to participants' to answer with their ideas about factors affected to the software quality risks. They were providing many answers and comments inadequate impact analysis, lack of traceability within requirements, development and testing, lack of peer and lead reviews and unit testing, no proper root cause analysis and actions (preventive and corrective), not adding more attention on continues improvement in process and product risks not identified in planning stage and not adding relevant testing types. However, these factors also affected to the software quality failure risks.

## 4.3 Summery

This study collected data from twenty-two participants through open-ended questionnaire circulated via email. Participants represented large scale, medium scale and small scale software companies in the Sri Lankan software industry. All categories of software quality professionals shared their experiences and comments about project quality failure risk. Thematic analysis approach was used for analyzing collected data. Major three themes from the data were identified, as the test coverage risk, planning, risk and lack of required resources risk.

## CHAPTER 5: CONCLUSION AND RECOMMENDATIONS

### 5.1 Introduction

This chapter provides the overall summery, summery of findings, limitations of the study and the recommendations based on the study findings.

### 5.2 Overall Summery

Software project failure and cancellation rates increase day by day (Emam & Koru, 2008). The majority of software cancellation happen even before any delivery occur to the end client. Due to the software project failures and cancellations, customers will have to bear high financial loses.

There are a number of reasons affecting software project failures. Rajkumar and Alagarsamy (2013) stated that lack of testing resources leads to poor quality. If product quality decreases, it will be a reason for project failure. Software quality failure is one of the major risks of a software project. Therefore, the research question for this study was 'What are the factors affecting the risk factors affected by the software quality failures? The main objectives of this study was to study the risk factors affecting the software quality and suggest some recommendations to avoid or minimize software project quality failure risks.

This study would help software quality managers, leads and responsible parties to understand how much testing; test planning and QA team skills impact software product quality. Emam & Koru (2008) stated that there is a 11% probability to have critical quality problems with software.

According to Schulmeyer (2007) software quality should have attributes and the satisfaction or degree of attainment of the attributes. The three components of Ould's approach are the business problem and its analysis, the risk and quality plan, and the project resource plan (Tsoukakas, J., 2001). Therefore, the proper testing, test planning and QA team impact the software quality risks were identified.

Furthermore, to investigate deep quality risk factor analysis, five questions for each risk factor were used. Circulate questionnaire among software company's quality assurance teams. There were 21 qualitative data responses collected from analysis, participants experience and comments. Qualitative data analysis was carried out by using the thematic approach. (Bruan & Clarke, 2006).

## 5.3 Summery of findings

Study sample size varied significantly in size, ranging from very small (21 employees) to very large (80,000 employees) (SLASSCOM, 2016) (Bannerman, 2008). The education background of participants represents their subject knowledge and project role displays their experience level. Company maturity is represented from company scale. Participants get different experiences while working on different types of projects and different sized teams. Participants were selected to represent all categories of information, the researcher expects to study.

Table 5.1: Study Profile

|  | Percentage (%) |
|---|---|
| **I. Education Background** |  |
| Postgraduate level | 55% |
| Degree level | 45% |
|  |  |
| **II. Participants Project Roll** |  |
| QA Manager | 14% |
| QA Lead | 36% |
| Senior QA Engineer | 27% |
| QA Engineer | 23% |
|  |  |
| **III. Company size/ scale** |  |
| Large | 23% |
| Medium | 63% |
| Small | 14% |
|  |  |
| **IV. Project type** |  |
| Development project | 77% |
| Maintenance project | 23% |
|  |  |
| **V. QA Team size** |  |
| Between 0 to 3 | 41% |
| Between 3 to 6 | 32% |
| Larger than 6 | 27% |

According to the data analysis, three theme's summaries were identified as follows,

Theme 1: Lack of testing leads to the risk of poor project quality

According to the experiences of the participants they agreed only with requirement clarity and clearly defined acceptance criteria, not doing enough unit testing and finally not doing regression testing force to quality failures.

Theme 2: Lack of proper planning leads to the risk of poor project quality

As of data analysis, not having proper formal test planning, initial test planning not being realistic, not following quality risk management, non-proper process and contingency action planning also lead to the risk of poor project quality.

Theme 3: Does not with proper QA Team resource lead to project quality risk.

According to the participants' comments following factors are also reasons to lack quality of software. The experienced and skilled employees move out from the company, not following proper QA process, and team members not having the risk management mentality.

### 5.4 Limitation of the study

There were few limitations of this study. This scope of the study is limited to software quality risk factor analysis. There are lots of project management study areas. This study selected project risk management study area and project scope narrowed down to quality risk management. As of Costa et al (2007), evaluating software project risks, consists of the key areas, analysis, design, coding, testing, planning, control, team, policies and structure, contract and client. With the help of the literature, three key areas out of ten are selected as testing, planning and team for this study.

This population for this study was quality assurance professionals in the software development industry. Here the data collection scope is limited to software firms in Colombo district Sri Lanka. Participant roles were limited to software quality assurance bodies. There were no project managers, technical team members and client's feedbacks or comments used for the analysis.

**5.5 Recommendation**

Quality Controlling was testing and it proved that the process can successfully produce the product, and then implement the proven process in operation (Chemuturi, 1950). They should be on inspection and testing binds with the cost of appraisal. Therefore, QA managers and leads are responsible for minimizing appraisal cost and improving quality of the product. For that they need to follow proper and standard methods for testing each and every type. Furthermore, they need to estimate the effort, time and budget correctly and execute them in an honest manner.

As a theory of the system test planning was how to get ready and organize for test execution. A test plan provides a framework, scope, details of resource needed, effort required, schedule of activities, and budget (Naik & Tripathy, 2008). However, Keogh (1994) stated that prevention costs arise in the course of preventing, investigating or reducing the risk of nonconformities or defects. Prevention costs may include, quality planning and QA team stability. According to Keogh's comment (1994), TQM program is essential to the message they communicate to all staff, for their active support of the QA managers, and for their decision making which affects the use of scarce resources. Recommendation for creating good and executable planning, QA managers always work with project manager and other team members. It should be a team work. QA managers always should listen to others as well and should prepare a collaborative test plan. Finally, QA leads are responsible for communicating each and every team member of the project and drive created plan correctly.

QA Team size impacts the delivered quality of the product, the development cost of the product, and the time to deliver the product (Naik & Tripathy, 2008). Furthermore, Naik and Tripathy (2008) states, organization's responsibility to define each and every team member's career path and responsibility level. Also, organization can setup performance based or experience based career paths criteria to move one role to another. Another important action is to conduct mentoring programs to each and every team member. Naik and Tripathy (2008) highlighted acknowledging and celebrating the group accomplishment are a powerful way to recognize the team effort and to keep the motivation and momentum afloat. Therefore, designing the team member's recognition system is very important.

**REFERENCES**

Abran, A., Nguyenkim, H. (1991) 'Analysis of maintenance work categories through measurement', IEEE Comput. Soc. Press, 1991, pp. 104-113.

Arachchilage, N. A. G. (2012). Security awareness of computer users: A game based learning approach. Ph.D. dissertation, Brunel University, School of Information Systems, Computing and Mathematics. http://bura.brunel.ac.uk/handle/2438/7620. Accessed 19 November 2016.

Arachchilage, N. A. G. (2015). User-Centred Security Education: A Game Design to Thwart Phishing Attacks. arXiv preprint arXiv:1511.03459.

Arachchilage, N. A. G., & Love, S. (2013). A game design framework for avoiding phishing attacks. Computers in Human Behavior, 29(3), 706-714.

Arachchilage, N. A. G., & Love, S. (2014). Security awareness of computer users: A phishing threat avoidance perspective. Computers in Human Behavior, 38, 304-312.

Arachchilage, N. A. G., & Martin, A. P. (2014). A Trust Domains Taxonomy for Securely Sharing Information: A Preliminary Investigation. In HAISA (pp. 53-68). arXiv preprint arXiv:1511.04541.

Arachchilage, N. A. G., Namiluko, C., & Martin, A. (2013). Developing a Trust Domain Taxonomy for Securely Sharing Information Among Others, International Journal for Information Security Research (IJISR), Volume 3, Issues 1 and 2, March/June 2013. arXiv preprint arXiv:1512.06307.

Arachchilage, N. A. G., Love, S., & Beznosov, K. (2016). Phishing threat avoidance behaviour: An empirical investigation. Computers in Human Behavior, *60*, 185-197.

Arachchilage N A G. Serious Games for Cyber Security Education. (2016). LAP Lambert Academic Publishing, 2016, eprint arXiv: 1610.09511. https://arxiv.org/abs/1610.09511 (accessed 15.11.2016).

Arachchilage, N. A. G., (2016). Serious games for cyber security education. Lambert Academic Publishing, pp. 1–244, ISBN-13: 978-3-659-85318-0. [arXiv preprint arXiv: 1610.09511]. Accessed 15 November 2016.

Arachchilage, N. A. G. (2013). Gaming for Security. ITNOW, 55(1), 32-33.

Arachchilage, N. A. G., Namiluko, C., & Martin, A. (2013, December). A taxonomy for securely sharing information among others in a trust domain. In Internet technology and secured transactions (ICITST), 2013 8th international conference for (pp. 296-304). IEEE.

A guide to the project management body of knowledge, 2013, 4th edn, Project management Institute, Inc, Pennsylvania, USA

Baccarini D., Salm G., Love P. E.D., (2004),"Management of risks in information technology projects", Industrial Management & Data Systems, Vol. 104 Iss 4 pp. 286 – 295.

Bahamdain, S. S. (2015) 'Open Source Software (OSS) Quality Assurance: A Survey Paper', Procedia Computer Science, 56(2015), pp. 459-464.

Bannerman, P. L. (2008) 'Risk and risk management in software projects: A reassessment', The Journal of Systems and Software, 81, pp. 2118–2133.

Benaroch, M., Lichtenstein, Y., Robinson, K. (2006) 'Real Options in Information Technology Risk Management: An Empirical Validation of Risk-Option Relationships', MIS Quarterly, 30(4), pp. 827-864.

Bengtsson, M. (2016) 'How to plan and perform a qualitative study using content analysis', NursingPlus Open, 2(2016), pp. 8-14.

Boehm, B. W. (1984) 'Software Engineering Economics', IEEE Transactions on Software Engineering, SE-10(1), pp. 4 - 21.

Boehm, B.W., Sullivan, K. (2000) 'Software Economics: A Roadmap', The Future of Software Engineering, 22nd International Conference on Software Engineering, pp. 321-343.

Braun, V., Clarke, V., (2006) 'Using thematic analysis in psychology', Qualitative Research in Psychology, 3(2), pp. 77-101.

British Standards Institution (1992) BS 7850: Part 1. Total Quality Management: Part 1. Guide to Management Principles, London: BSI.

Burch, J. G., Grupe, F. H. (1993) 'Improved software maintenance management', Journal Information Systems Management, 10(1), pp. 24-32.

Chappell, D. (2013) The three aspects of software quality: functional, structural, and process, Microsoft Corporation.

Charette, R. N., 2005. Why software fails [software failure]. IEEE Spectrum, [Online]. 42, 42-49. Available at: http://ieeexplore.ieee.org/document/1502528/ [Accessed 11 September 2016].

Chemuturi, M (1950) Mastering Software Quality Assurance: Best Practices, Tools and Techniques for software developers, 1st edn., Fort Lauderdale, Florida: J. Ross Publishing Inc.

Cherapau, I., Muslukhov, I., Asanka, N., & Beznosov, K. (2015, July). On the impact of touch id on iphone passcodes. In *Eleventh Symposium On Usable Privacy and Security (SOUPS 2015)* (pp. 257-276). USENIX Association.

Cooper, D. R., Schindler, P.,Sharma, J. K., 2012. Business Research Methods. 11th ed. New Delhi: Tata McGraw Hill Education Private Limited.

Cope, R. F., Folse, R. O., Cope, R. F. (1999) 'Quality control for software warranties: a conceptual and economic perspective', Management Research News, 22(7), pp. 30 - 40.

Costa, H. R., Barros, M. O., Travassos, G. H. (2007) 'Evaluating software project portfolio risks', Journal of Systems and Software, 80(2007), pp. 16-31.

Dey, I. (1993) Qualitative Data Analysis, 1st edn., 11, New Fetter Lane, London EC4P 4EE: Routledge.

Donath, C., Winkler, A., Graessel, E., & Luttenberger, K. (2011) 'Day care for dementia patients from a family caregiver's point of view: A questionnaire study on expected quality and predictors of utilisation - Part II', BMC Health Services Research, 2011(11), pp. 1-7.

Downe-Wamboldt, B (1992) 'Content analysis: Method, applications, and issues', Health Care for Women International, 13(1992), pp. 313-321.

Emam, K. E., Koru, A. G. (2008) 'A Replicated Survey of IT Software Project Failures', IEEE Software, 25(5), pp. 84-90.

Fridlund, B., Hildingh, C. (2000) 'Health and qualitative analysis methods In: B. Fridlund, & C. Hildingh (Eds.)', Qualitative research, methods in the service of health, Lund: Studentlitteratur(2000), pp. 13-25.

Gacek, C., Arief, B. (2004) 'The many meanings of open source', IEEE Software, 21(1), pp. 34 - 40.

Galin, D. (2009) Software Quality Assurance: From Theory to Implementation, 1st edn., New Delhi, India: Dorling Kindersley (India) Pvt. Ltd.

Graham, D., Veenendaal, E. V., Evans, I. (2008) Foundations of Software Testing: ISTQB Certification, 1st edn., London: Cengage Learning EMEA

Hameed, M. A., & Arachchilage, N. A. G. (2016). A Model for the Adoption Process of Information System Security Innovations in Organisations: A Theoretical Perspective. 27th Australasian Conference on Information Systems (ACIS), University of Wollongong, Australia

Hamill, P (2005) Unit Test Frameworks: Tools for High-Quality Software Development, 1st edn., Sebastopol, CA95472: O'Reilly Media, Inc.

Hammond, J.L., Hammond, B. (1917) The town labourer, 1760-1832; the new civilization.archive.org [Online]. Available at: https://archive.org/details/townlabourer00hammuoft (Accessed: 15th October 2016).

Iversen, J. H., Mathiassen, L., Nielsen, P. A. (2004) 'Managing Risk in Software Process Improvement: An Action Research Approach', MIS Quarterly, 28(3), pp. 395-433.

Jayawarna, S., Fonseka, A. (2011). Factors Affecting Product Quality in the Software Development Industry of Sri Lanka. Sri Lankan Journal of Management. 16, pp.134-137.

Jovanovic, V., Shoemaker, D. (1997) 'ISO 9001 standard and software quality improvement', Benchmarking for Quality Management & Technology, 4(2), pp. 148 - 159.

Keogh, W. (1994) 'The Role of the Quality Assurance Professional in Determining Quality Costs', Managerial Auditing Journal, 9(4), pp. 23 - 32.

Knox, S. T. (1993) 'Modeling the Cost of Software Quality', Digital Technical Journal, 5(4), pp. 9-17.

Koelsch, G (2016) Requirements Writing for System Engineering, 1st edn., Verginia, USA: Apress.

Kumar, C, Yadav, D. K (2015) 'A Probabilistic Software Risk Assessment and Estimation Model for Software Projects', Procedia Computer Science, 54(2015), pp. 353 – 361.

Kwak, Y. H., Stoddard, J. (2004) 'Project risk management: lessons learned from software development environment', Technovation, 24(2004), pp. 915–920.

List some disadvantages of SAP. . 2016. List some disadvantages of SAP. . [ONLINE] Available at:http://www.allinterview.com/showanswers/100972/list-some-disadvantages-of-sap.html. [Accessed 15 October 2016].

López, C., Salmeron, J. L., (2012) 'Monitoring Software Maintenance Project Risks',Procedia Technology, 5(2012), pp. 363-368.

Maidasani, D. (2007) Software Testing, 1st edn., New Delhi: Laxmi Publications Pvt. Ltd.

March, J. G., and Shapira, Z. (1987) 'Managerial Perspectives on Risk and Risk Taking', Management Science, 33(11), pp. 1404-1418.

Mc Connell, S. (1998) Software project survival guide, 1st edn., A Division of Microsoft Corporation, One Microsoft Way, Redmond, Washington 98052-6399: Microsoft Press.

Mooney, C, 2011. The industrial revolution investigate how science and technology changed the world with 25 Projects. 1st ed. Vermont, USA: Nomad Press.

Mortiboys, R.J. and Oakland, J.S. (1991) Total Quality Management and Effective Leadership, London: Department of Trade and Industry.

Naik, K., Tripathy, P. (2008) Software Testing and Quality Assurance: Theory and Practice, 1st edn., Hobokan, New Jersey: John Wiley & Sons.

Nielsen,J., Mack, R. L., Bergendorff, K. H., Grischkowsky, N. L. (1986) 'Integrated software usage in the professional work environment: evidence from questionnaires and interviews', CHI '86 Proceedings of the SIGCHI Conference on Human Factors in Computing Systems, 17(4), pp. 162-167.

O'Brochta, Michael. (2002). "Project Success – What Are the Criteria and Whose Opinion Counts?", Proceedings of the Project Management Institute Annual Seminars & Symposiums, October 3 – 10, 2002,  San Antonio, TX.

Otte, T., Moreton, R., Knoell, H. D. (2008) 'Applied Quality Assurance Methods under the Open Source Development Model', 2008 32nd Annual IEEE International

Computer Software and Applications Conference, July 28 2008-Aug. 1 2008, pp. 1247-1252.

Ould, M. A. (1999) Managing software quality and business risk, 1st edn., Hoboken, New Jersey, United States: John Wiley & Sons.

Patel, A, Mohanan, PP, Prabhakaran, D & Huffman, MD 2016, 'Pre-hospital acute coronary syndrome care in Kerala, India: A qualitative analysis' Indian Heart Journal. DOI: 10.1016/j.ihj.2016.07.011

Pinto, J.K., Mantel, S.J. (1990) 'The Causes of Project Failure', IEEE Software, 37(7), pp. 269 - 276.

Pritchard, C.L. (1997), Risk Management – Concepts and Guidance, ESI International, Arlington, VA.

Rajkumar, G., Alagarsamy, K. (2013) 'The most common factors for the failure of software Development project', The International Journal of Computer Science & Applications, 1(11), pp. 74-77.

Raju, R. S., Parthasarathy, A. (2009) MANAGEMENT: Text and Cases, 2nd edn., New Delhi: PHI Lerning Pvt. Ltd.

Regan, G. O. (2002) A Practical Approach to Software Quality, 1st edn., New York: Springer Science.

Schmidt, R., Lyytinen, K., Keil, M., Cule, P. (2001) 'Identifying Software Project Risks: An International Delphi Study', Journal of Management Information Systems, 17(4), pp. 5-36.

Schulmeyer, G. G. (2007) Handbook of Software Quality Assurance, 4th edn., 685, Canton Street, Norwood, MA 02062: Artech House Inc.

Shalloway, A.. 2016. Net Objectives. [ONLINE] Available at: http://www.netobjectives.com/blogs/cause-poor-software-quality. [Accessed 10 January 2017].

Software Enineering Institute (2010) CMMI for Development, Version 1.3, ESC/XPK, 5, Eglin Street, Hanscom AFB, MA 01731-2100: SEI Administrative Agent.

Sri Lanka Association of Software and Service Companies (2016) Sri Lankan IT/BPM Industry 2016 Review, PRICEWATERHOUSECOOPERS, Sri Lanka: A SLASSCOM publication.

Standard Coordinating Committee of the Computer Society of the IEEE (1990) IEEE Standard Glossary of Software Engineering Terminology, New York, USA: The Institute of Electrical and Electronic Engineers.

Tsoukakas, J. (2001) 'Managing Software Quality and Business Risk', *JSTOR,* 31(6), pp. 123-125 [Online]. Available at: *http://www.jstor.org/stable/25062760* (Accessed: 27-08-2016).

Whittaker, B. (1999) 'What went wrong? Unsuccessful information technology projects', Information Management and Computer Security, 7, pp. 23-30.

Williams, T. (1995) 'A classified bibliography of recent research relating to risk management', European Journal of Operational Research, 85(1), pp. 18-38.

Zhi, H. (1994), "Risk management for overseas construction projects", International Journal of Project Management, Vol. 13 No. 3, pp. 231-7.

Zmud, R. W. (1980) 'Management of Large Software Development Efforts', MIS Quarterly, 4(2), pp. 45-55.

**APPENDIX I – SOFTWARE QUALITY RISK QUESTIONNAIRE**

The following questions represent categories of software project quality risk factors analysis.

**Background of Participant:**

1. What is your Education background?

2. What is your current project roll?

3. What is your Current Company?

4. What is your company size/ scale?

5. What is your Project type?

6. What is your QA team size?

7. Do your project have risk of failure?

**Testing related risks:**

1. Have acceptance criteria been agreed to for all requirements?

2. Has sufficient unit testing been specified?

3. Has adequate time been allocated for product integration and testing?

4. Does hardware and software instrumentation facilitate testing?

5. Is regression testing performed?

**Quality planning related risks:**

6. Is there a formal testing plan?

7. Is there an effective QA risk management process?

8. Are there contingency plans for known QA risks?

9. Is the software quality assurance function adequately staffed on this project

10. Is the QA schedule realistic?

**Quality Assurance Team skills related risks:**

11. Do people get trained in skills required for this project?

12. Is there any problem keeping the skilled people you need?

13. Do people work effectively towards common goals?

14. Are all staff levels oriented toward quality procedures?

15. Is risk management mentality part of the team culture?

**APPENDIX I I– SOFTWARE QUALITY RISK INTERVIEW QUESTION**

**Background of Participant:**

1. What is your Education background?

2. What is your current project roll?

3. What is your Current Company?

4. What is your company size/ scale?

5. What is your Project type?

6. What is your QA team size?

7. Do your project have risk of failure?

8. What are the Quality Risk factors?